\documentclass[useAMS,usenatbib]{mn2e}
\usepackage{graphicx,fleqn,natbib,amssymb}
\DeclareGraphicsExtensions{.eps,.cps,.ps,.eps.gz,.ps.gz,.eps.Z}
\usepackage{multirow}

\title[$\rm log\;L_{200\rm{s}}-\alpha_{>200\rm{s}}$ correlation]{A correlation between intrinsic brightness and average decay rate of {\it Swift} UVOT GRB optical/UV light curves}

\author[Oates et al.]{S. R. Oates$^{1}$, M. J. Page$^{1}$, M. De Pasquale$^{2}$, P. Schady$^{3}$, A. A. Breeveld$^{1}$ 
\newauthor S. T. Holland$^{4}$, N. P. M. Kuin$^{1}$, F. E. Marshall$^{5}$  \\
$^{1}$ Mullard Space Science Laboratory, University College London, Holmbury St. Mary, Dorking Surrey, RH5 6NT, UK; sro@mssl.ucl.ac.uk \\
$^{2}$ University of Nevada, Las Vegas, Department of Physics \& Astronomy, Box 454002, 4505 Maryland Parkway, Lax Vegas, NV 89154-4002\\ 
$^{3}$ Max-Planck Institut f\"{u}r Extraterrestrische Physik, Giessenbachstra\ss e 1, 85748 Garching, Germany\\
$^{4}$ Space Telescope Science Center, 3700 San Martin Dr., Baltimore, MD 21218, USA\\
$^{5}$ Astrophysics Science Division, Code 660.1, NASA Goddard Space Flight Centre, 8800 Greenbelt Road, Greenbelt, Maryland 20771, USA \\}

\begin{document}
\date{\vspace{-5ex}}


\maketitle

\label{firstpage}

\begin{abstract} 
We examine a sample of 48 {\it Swift}/UVOT long Gamma-ray Burst light curves and find a correlation between the logarithmic luminosity 
at 200s and average decay rate determined from 200s onwards, with a Spearman rank coefficient of -0.58 at a significance of 99.998\% 
(4.2$\sigma$). We discuss the causes of the $\rm log\;L_{200\rm{s}}-\alpha_{>200\rm{s}}$ correlation, finding it to be an intrinsic 
property of long GRBs, and not resulting from the selection criteria. We find two ways to produce the correlation. One possibility is 
that there is some property of the central engine, outflow or external medium that affects the rate of energy release so that the 
bright afterglows release their energy more quickly and decay faster than the fainter afterglows. Alternatively, the correlation may 
be produced by variation of the observers viewing angle, with observers at large viewing angles observing fainter and slower decaying 
light curves.

\end{abstract}


\section{Introduction}
\label{intro}

Gamma-ray bursts (GRBs) are intense flashes of gamma-rays that are usually accompanied by an afterglow, longer lived emission that 
may be detected at X-ray to radio wavelengths. Our understanding of GRB X-ray and optical/UV afterglows was revolutionised by 
the launch of {\it Swift}, a satellite dedicated to the detection of GRBs and observation of their $\gamma$-ray, X-ray and 
optical/UV emission \citep{geh04}. The best studied subclass of GRB are the long GRBs (LGRBs), which have observed prompt $\gamma$-ray 
emission durations of $\gtrsim 2$s \citep{kou93}. Several studies have investigated the X-ray emission of LGRBs, using large 
samples, to identify characteristic temporal and spectral behaviours \citep[i.e][]{nousek,eva07,eva09}. Similar investigations 
have been performed at optical/IR wavelengths \citep[i.e][]{mel08,ryk09,oates09}, but these have tended to have much smaller 
samples, due to the lower detection rate of the optical emission in comparison to the X-rays (the detection rate for {\it Swift}'s 
X-ray and ultra-violet optical telescopes are $\sim96\%$ \citep{bur08} and $\sim29\%$ \citep{rom09}, respectively for 
LGRBs). The low detection rate is generally attributed to extinction due to high levels of dust in the host \citep{fyn01}, and/or to high redshift at 
which the optical/UV emission will be absorbed by neutral hydrogen along the line of sight \citep{gro98}; see also \cite{gre11} for 
a recent study on the cause of optically dark GRBs. However, {\it Swift} has now observed over 600 GRBs and we are now in a position 
to collate a large number of well-sampled IR/optical/UV LGRB afterglows from which we can draw a representative picture of 
their behaviour and collective properties \citep[see recent papers by][]{kan10,li12}. 

In \cite{oates09}, we performed a statistical investigation of 26 optical/UV LGRB afterglows, finding a correlation between the 
observed $v$-band magnitude at 400s and the average UVOT light curve decay rate determined from 500s. We also tested for an 
equivalent rest frame correlation, but, due to the small sample size, this could not be confirmed or excluded. Here we use a 
larger sample of 48 high quality LGRB UVOT light curves to re-examine if there is a correlation between optical/UV afterglow 
intrinsic brightness and light curve decay rate. 

This paper is organized as follows. In \S~\ref{reduction} we describe the data reduction and analysis. The main results are 
presented in \S~\ref{results}. Discussion and conclusions follow in \S~\ref{discussion} and \ref{conclusions}, respectively. 
All uncertainties throughout this paper are quoted at 1$\sigma$. The temporal and spectral indices, $\alpha$ and $\beta$, are 
given by the expression $F(t,\nu)\propto t^{\alpha}\nu^{\beta}$.

\begin{figure}
\includegraphics[angle=-90,scale=0.30]{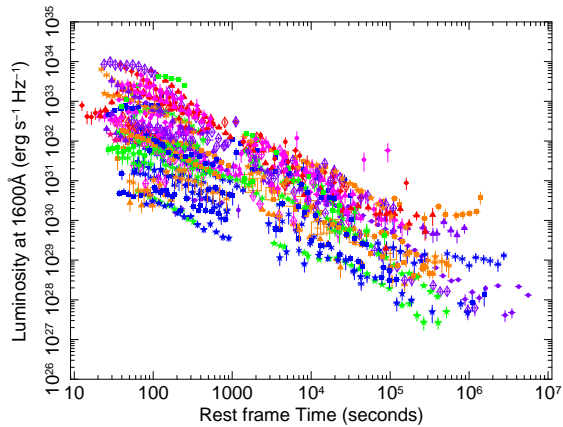}
\caption{Optical luminosity light curves of 56 GRBs at restframe 1600\AA. For clarity, 3 $\sigma$ upper 
limits are not included.}
\label{lightcurves}
\end{figure}

\section{Data Reduction and Analysis}
\label{reduction}

\subsection{SAMPLE}
Our sample began with 69 LGRBs from the second {\it Swift} UVOT GRB afterglow catalogue \citep{rom}, which 
were observed between April 2005 and December 2010. The sample was selected using the same criteria as \cite{oates09}: the 
optical/UV light curves must have a peak UVOT $v$-band magnitude of $\leq$17.89 (equivalent to 1 ct $\rm s^{-1}$), UVOT 
must observe within the first 400s until at least $10^5$s after the BAT trigger and the colour of the afterglows must 
not evolve significantly with time, meaning that at no stage should the light curve from a single filter significantly 
deviate from any other filter light curve when normalized to the $v$ filter. Only GRBs 060218 and 100814A were 
excluded as we considered them to have strong colour evolution. These criteria ensure that a high signal-to-noise 
(SN) light curve, covering both early and late times, can be constructed from the UVOT multi-filter observations.

In order to obtain the best SN light curve for each GRB, a single light curve was constructed from the multi-filter light 
curves, following the method in \cite{oates09}. The main steps are to normalize the multi-filter light 
curves to the $v$ filter and then to group them together using a binsize of $\Delta t/t=0.2$. Following \cite{oates09}, 
 for each GRB, we take the onset of the prompt $\gamma$-ray emission as the start time of the UVOT light curve. Since the 
BAT may not trigger at the start of the prompt emission, we take the start time of the $T_{90}$ parameter\footnotemark to 
be the start time for the UVOT light curve.

\footnotetext[1]{The $T_{90}$ parameter is determined from the gamma-ray event data for each GRB, by the BAT processing 
script. The results of the processing are publicly available and are provided for each trigger 
at http://gcn.gsfc.nasa.gov/swift\_gnd\_ana.html.}

\subsection{Luminosity Light curve}
Luminosity light curves were produced for all GRBs with spectroscopic or photometric redshifts in the literature 
and for which host E(B-V) values could be determined. For a further 3 GRBs we were able to derive photometric 
redshifts from joint XRT-UVOT Spectral Energy Distributions (SEDs), following the method of \cite{kru11}: 
$z=1.2\pm0.1$ for GRB~060510A, $z=3.1\pm0.1$ for GRB~090401B and $z=1.85\pm0.10$ for GRB~100805A. For 13 GRBs, 
no redshift was available nor could we derive a photometric redshift.

For each of the 56 GRBs with host E(B-V) and redshift measurements, we converted the single filter count rate light 
curve to luminosity at a common rest frame wavelength. In order to select the common wavelength and determine 
the resulting k-correction factor for each light curve, an SED was computed for each GRB following the 
methodology in \cite{oates09} and using the count rate-to-flux conversion factors given in \cite{bre11}. 
The common rest frame wavelength was selected to maximise the number of GRBs with SEDs that covered this 
wavelength and to be relatively unaffected by host extinction. As was found in \cite{oates09}, the wavelength that 
best satisfies these conditions is $1600${\AA}. For each GRB, the k-correction factor, $k$, was taken as the flux 
density at the wavelength that corresponds to 1600{\AA} in the rest frame, $F_{1600}$, divided by the flux density at 
the observed central wavelength of the $v$ filter (5402{\AA}), $F_{v}$, which was multiplied by $(1+z)$, where $z$ 
is the redshift of the GRB such that $k=(F_{1600}/(F_{v}*(1+z)))$. For those GRBs with SEDs not covering 1600{\AA}, an 
average $k$ value was determined from the other GRBs in the sample, which have SEDs covering both 1600{\AA} and 
the $v$ filter rest frame wavelength.

To convert to luminosity, the count rate light curves were initially corrected for Galactic extinction and 
then converted into flux per unit frequency. These flux densities were then converted to luminosity 
at 1600{\AA} using: 
\begin {equation}L(1600)=4\pi D_L^2 F_{v}k\label{luminsoity}\end{equation} 
where $L(1600)$ is the luminosity at a 1600{\AA} and $D_L$ is the luminosity distance. Finally, the luminosity 
light curves were corrected for host extinction, which was calculated from the host $A_{v}$ values reported 
in \cite{sch10}. For those not reported in \cite{sch10}, we used the same method to determine the host $A_{v}$.

\section{Results}
\label{results}
In Figure \ref{lightcurves}, we show the luminosity light curves at 1600 \AA, in units of $\rm erg\;s^{-1}\;Hz^{-1}$.
The light curves are clustered in a single group, having the largest range in luminosity at the earliest epochs, 
which becomes narrower as the light curves decay. This characteristic is also seen in the light curves given in 
\cite{kan10}. The narrowing of the range in luminosity with time suggests that the most luminous GRBs decay the 
quickest and the less luminous GRBs decay more slowly. In the next section, we determine if there is a statistically 
significant correlation between the brightness and decay rate and quantify this correlation. 

\subsection{Luminosity Decay Correlation}
\label{restdecaycorr}
\begin{figure}
\includegraphics[angle=0,scale=0.48]{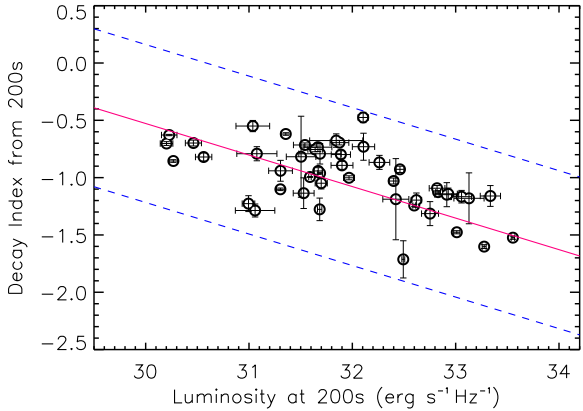}
\caption{Average decay index determined from the luminosity light curves after 200s versus luminosity at 200s. The red 
solid line represents the best fit regression and the blue dashed line represents the 3$\sigma$ deviation.}
\label{correlation}
\end{figure}

In order to test for a correlation we determined the intrinsic brightness at 200s and the average decay rate 
determined from 200s onwards. To determine the intrinsic brightness, we used the IDL interpolation function, {\sc interpol}, 
on the data between 100 and 2000s and interpolated the logarithmic luminosity at 200s, $\rm log\;L_{200\rm{s}}$. For the 
average decay rate, $\alpha_{>200\rm{s}}$, using error weighted least squares we fit a single power law to the light curves 
from 200s onwards. A restframe time of 200s was chosen as by this time all the light curves in the sample 
have observations. For six GRBs, there were too few data points between 100 and 2000s to constrain the luminosity 
at 200s and we excluded two further GRBs for which the error on the decay index was greater than 3 times the sample 
mean decay error. The resulting values for $\rm log\;L_{200\rm{s}}$ and $\alpha_{>200\rm{s}}$ are shown in 
Fig. \ref{correlation}. A Spearman rank test of $\rm log\;L_{200\rm{s}}$ against $\alpha_{>200\rm{s}}$, gives a correlation 
coefficient of -0.58 at a significance of 99.998\% (4.2$\sigma$). This indicates that $\rm log\;L_{200\rm{s}}$ and 
$\alpha_{>200\rm{s}}$ are statistically correlated and confirms that luminous optical/UV afterglows decay more quickly 
than less luminous ones.

\begin{figure}
\includegraphics[angle=0,scale=0.45]{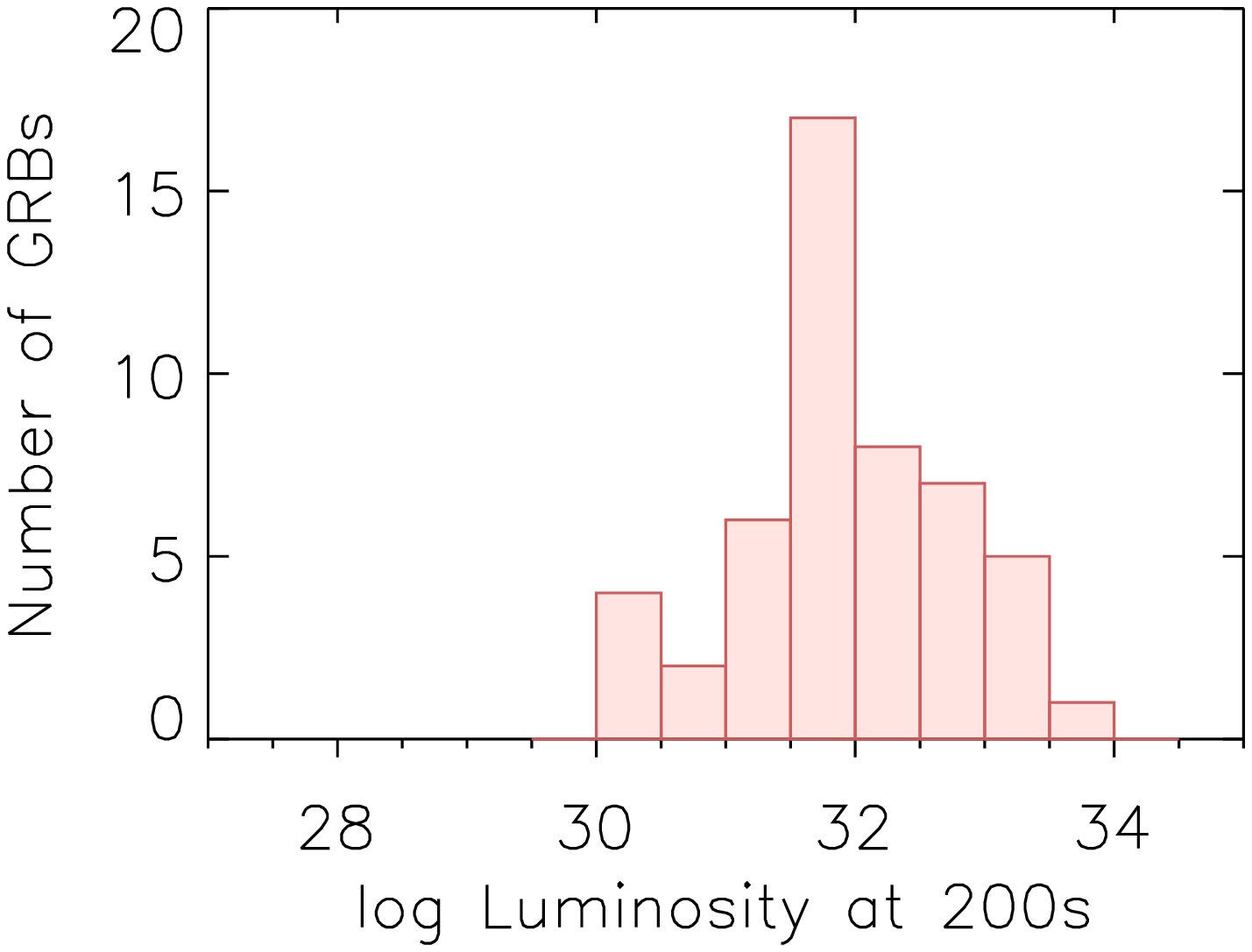}
\includegraphics[angle=0,scale=0.45]{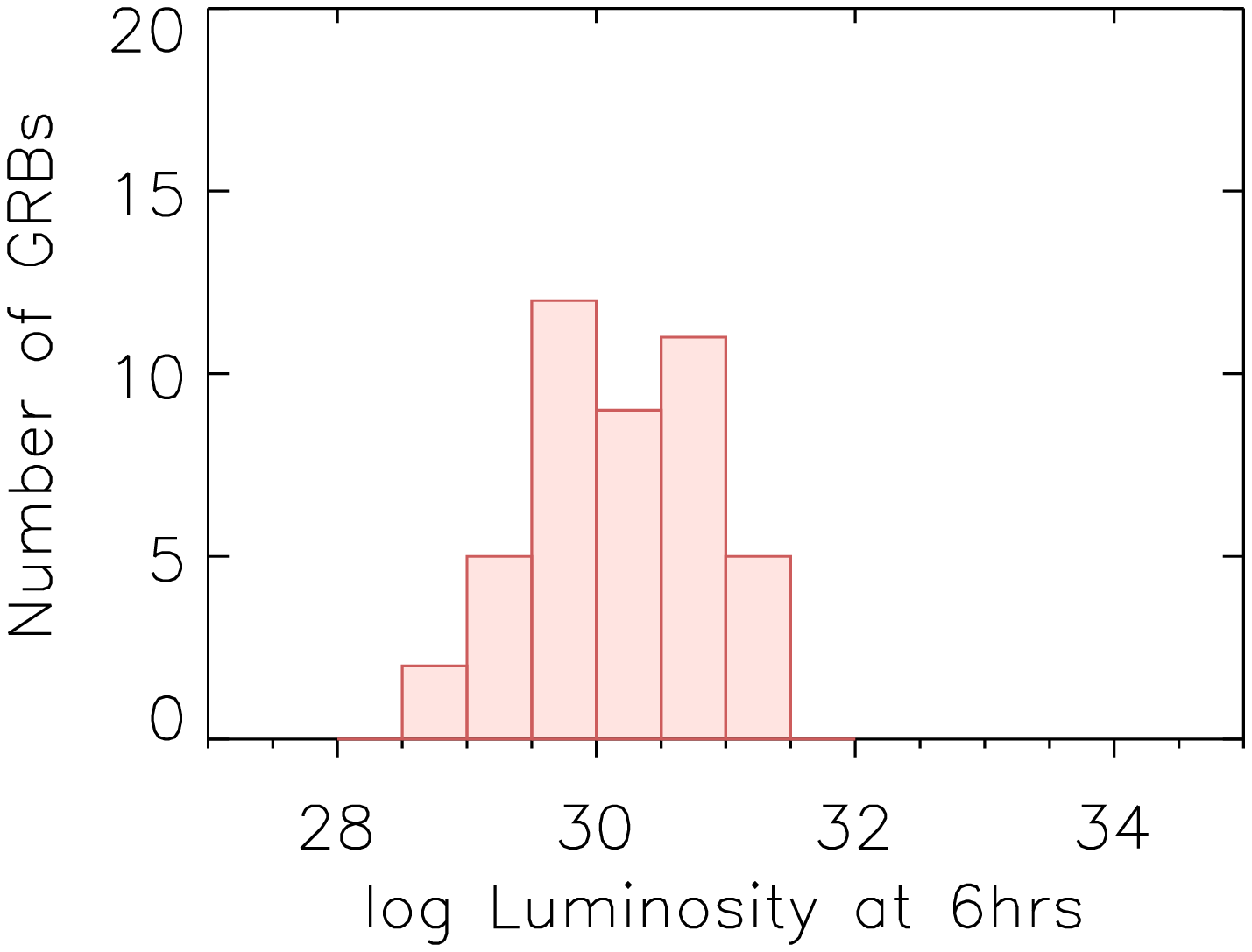}
\caption{Distribution of logarithmic luminosities at rest frame 200s (top) and 6hrs (bottom).}
\label{lum_dist}
\end{figure}

In Fig. \ref{lum_dist}, we show the distributions of the logarithmic luminosity at 200s and 6 hours, $\rm log\;L_{6hrs}$. 
If the brighter optical/UV afterglows decay more quickly than the fainter ones then at late epochs the luminosity 
distribution must become narrower and the correlation should become weaker and/or insignificant. We observe both of 
these effects in our sample. In Fig. \ref{lum_dist}, the 6hr luminosity distribution has a standard deviation of 0.69 
which is smaller than the standard deviation of the 200s luminosity distribution, which is 0.84 \citep[see also][]{kan10}, 
and, a Spearman rank test of $\rm log\;L_{6hrs}$ and $\alpha_{>200\rm{s}}$ indicates no significant correlation with 
a coefficient of 0.01 and a significance of 8\%.

In order to quantify the relationship between $\rm log\;L_{200\rm{s}}$ and $\alpha_{>200\rm{s}}$, we performed a linear 
regression using the IDL routine, {\sc fitexy}, which takes into account the errors on both parameters. This 
analysis provides a linear relationship of $\rm log\;L_{200\rm{s}}=(-3.636\pm0.004)\alpha+(28.08\pm0.13)$, given as 
a solid red line in the left panel of Fig. \ref{correlation}. All GRBs reside within $3\times$ the rms deviation, indicated 
by blue dotted lines. 

We looked to see if the $\rm log\;L_{200\rm{s}}-\alpha_{>200\rm{s}}$ correlation was due to both parameters 
being related to redshift, $z$. A Spearman rank correlation between these parameters gives: a 
coefficient of 0.62 at a significance of $4.7\sigma$ for ${\rm log\;L_{200\rm{s}}}-z$, and a coefficient of -0.33 at a significance of 
$98\%$ for $\alpha_{>200{\rm s}}-z$. The correlation between ${\rm log\;L_{200\rm{s}}}$ and redshift is expected as we are able 
only to detect GRBs at high redshift if they are very luminous. The correlation between $\alpha_{>200\rm{s}}$ and redshift is 
much weaker than the correlation between $\rm log\;L_{200\rm{s}}$ and $\alpha_{>200\rm{s}}$ suggesting that the 
$\rm L_{200\rm{s}}-\alpha_{>200\rm{s}}$ correlation is not due to the implicit correlation between these two parameters and 
redshift. To confirm this, we calculate the partial Spearman rank correlation, which measures the degree of correlation 
between two parameters, $\rm L_{200\rm{s}}$ and $\alpha_{>200\rm{s}}$, excluding the effect of a third, in this case 
redshift \citep[see][for further details]{ken79}. Using this method we obtain a correlation coefficient of -0.50 
with a confidence of $99.97\%$ (3.5$\sigma$), only a small reduction in the correlation coefficient from the standard Spearman 
rank correlation. This indicates that the correlation between $\rm L_{200\rm{s}}$ and $\alpha_{>200\rm{s}}$ is not a 
result of the implicit correlation between these two parameters and redshift.

The $\rm log\;L_{200\rm{s}}-\alpha_{>200\rm{s}}$ correlation could be due to chance or be a result of selecting the sample 
by their observed frame properties, specifically the exclusion of LGRB optical afterglows that were fainter than 17.89 in 
UVOT $v$-band and/or were not observed within 400s after the trigger. In order to eliminate these possibilities, we performed a 
Monte Carlo simulation. For each of a total of $10^6$ trials, we simulated a distribution of 48 pairs of $\rm L_{200\rm{s}}-\alpha_{>200\rm{s}}$ 
data points selected at random from linear distributions of $\rm L_{200\rm{s}}$ and $\alpha_{>200\rm{s}}$, which have the same ranges as the 
observed sample. For each pair of data points, we produced a synthetic observed frame light curve using $\rm L_{200\rm{s}}$ and 
$\alpha_{>200\rm{s}}$ and randomized values for redshift, host and Galactic extinction and $\beta$; where, $\beta$, was used to determine 
the k-correction factor \citep[$k=(1+z)^{1+\beta}$;][]{lam00}. Similar to the simulated $\rm L_{200\rm{s}}$ and $\alpha_{>200\rm{s}}$ points, 
these parameters were sampled randomly from linear distributions, which have the same range as the observed light curves. 
In order to simulate the time it takes for {\it Swift} to point its narrow field instruments at the GRB location, we selected 
at random an observed frame light curve from our sample and used the time sampling of this light curve, i.e photometry times 
and durations, as the time sampling for the simulated observed frame light curve. We then determined if the simulated observed 
frame light curve met our selection criteria. For those that did not meet the selection criteria, we discarded the corresponding 
$\rm L_{200\rm{s}}-\alpha_{>200\rm{s}}$ data point and drew a new pair of values until the selection criteria were met. Once 48 
pairs of $\rm L_{200\rm{s}}$ and $\alpha_{>200\rm{s}}$ had been found with synthetic observed frame light curves that met the 
selection criteria, a Spearman rank correlation was performed on the simulated $\rm L_{200\rm{s}}-\alpha_{>200\rm{s}}$ distribution.

Of the $10^6$ trials, only 34 have a correlation coefficient equal to or indicating a stronger correlation than the 
real $\rm L_{200\rm{s}}-\alpha_{>200\rm{s}}$ distribution. This indicates that, at $4.1\sigma$ confidence, the 
$\rm L_{200\rm{s}}-\alpha_{>200\rm{s}}$ correlation is not due to our selection criteria nor does it occur by 
chance and therefore implies that the $\rm L_{200\rm{s}}-\alpha_{>200\rm{s}}$ correlation is intrinsic to LGRBs. 

\begin{figure}
\includegraphics[angle=0,scale=0.48]{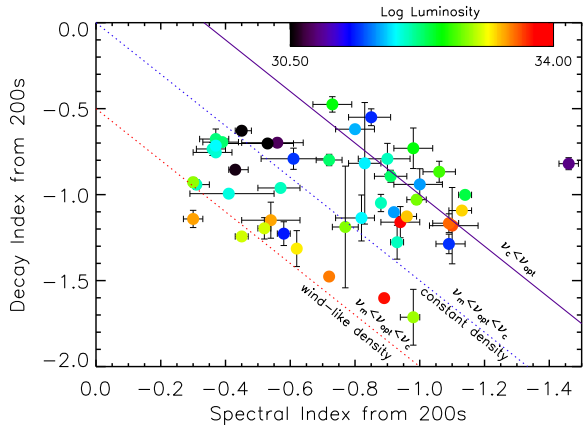}
\caption{Optical/UV temporal and spectral indices for the sample of 48 GRBs. The lines represent 3 closure relations 
and a colour scale is used to display the range in luminosity at 200s, $\rm L_{200\rm{s}}$. }
\label{beta}
\end{figure}

\section{Discussion}
\label{discussion}
Using 48 optical/UV light curves, we have found a significant correlation between the intrinsic brightness 
and average decay rate of LGRB optical/UV afterglows. A similar correlation was found by \cite{pan08}, who 
found for those GRBs with an observed rise in the optical light curves, there was a correlation between peak 
flux and post-peak power-law decay rate. The correlation reported by \cite{pan08} is limited to only those GRBs 
which are observed to rise.  Since the majority of rising afterglows cease rising by $\sim400$s \citep{oates09}, 
which is typically earlier than restframe 200s (i.e $\sim400s/(1+z)<200$s), the $\rm log\;L_{200\rm{s}}-\alpha_{>200\rm{s}}$ 
correlation determined in this work can be applied to optical afterglows with all types of early behaviour. Furthermore, 
as the UVOT typically begins observing within the first $100$s after the BAT trigger, the majority of GRBs observed 
by {\it Swift}, with a detected optical counterpart, can be included in the correlation reported in this paper. In 
the following, we shall examine the possible ways to produce a correlation between $\rm log\;L_{200\rm{s}}$ and $\alpha_{>200\rm{s}}$. 


\subsection{Is the correlation predicted by the standard afterglow model?}

\subsubsection{The basic model}
The $\rm log\;L_{200\rm{s}}-\alpha_{>200\rm{s}}$ correlation may be a natural result of the jet interacting with 
the external medium, producing synchrotron emission. Here we assume an isotropic, collimated outflow which is not
energy injected. In this model, the luminosity is related to $\alpha$ and $\beta$ by $L_\nu \propto F_\nu\propto t^{\alpha}\nu^{\beta}$, 
and $\alpha$ and $\beta$ are related linearly by the closure relations \cite[e.g][]{sar98}, defined by the 
external medium density profile and the ordering of the synchrotron frequencies. For the GRBs in this sample, due to 
our selection criteria requiring no colour evolution, we may conclude that $\beta$ does not 
vary during the course of our observations. We now examine two scenarios to determine if the correlation 
is a result of the basic afterglow model.

In the simplest scenario, all optical afterglows arise from a single closure relation, in which $\alpha$ is not a fixed value.
In this scenario, $\alpha$ and $\beta$ are related linearly and therefore a correlation between 
$\rm log\;L_{200\rm{s}}$ and $\beta$ should be expected. However, only a weak correlation is observed between $\beta$ and 
$\alpha$ and there is no evidence for a correlation between $\beta$ and $\rm log\;L_{200\rm{s}}$ (see Fig \ref{beta}), with 
Spearman rank coefficients of 0.26 (92\%) and -0.15 (68\%), respectively. This scenario 
is therefore not likely the cause of the $\rm log\;L_{200\rm{s}}-\alpha_{>200\rm{s}}$ correlation.

However, we do not expect all optical afterglows to be on the same phase of the synchrotron spectrum. In 
the second scenario, we assume that the $\rm log\;L_{200\rm{s}}-\alpha_{>200\rm{s}}$ correlation is a result 
of more than one closure relation, which requires the use of multiple spectral segments and different density 
profiles. The spectral segments most commonly found to satisfy optical afterglow production are: 
$\nu_m<\nu_{opt}<\nu_c$ and $\nu_c<\nu_{opt}$, where $\nu_c$ is the cooling frequency and 
$\nu_m$ is the peak frequency. The $\nu_m<\nu_{opt}<\nu_c$ spectral segment requires different closure 
relations for when the ejecta interacts with a constant density medium and a wind-like density medium. 
The spectral segment, $\nu_c<\nu_{opt}$, is independent of the external medium density profile. The 
expected relations between $\alpha$ and $\beta$ for these scenarios are indicated by lines in 
Fig. \ref{beta}. If the $\rm log\;L_{200\rm{s}}-\alpha_{>200\rm{s}}$ correlation was produced by the optical 
afterglows resulting from multiple closure relations, in Fig. \ref{beta} we would observe the $\alpha$ and $\beta$ 
data points with similar luminosities to cluster around a given closure relation. In Fig. \ref{beta}, the data points 
do not appear to display this trend and therefore we find it unlikely that the basic standard model is causing 
the $\rm log\;L_{200\rm{s}}-\alpha_{>200\rm{s}}$ correlation.

\subsubsection{Complex afterglow model}
The afterglow model is likely to be more complex than we previously assumed and there may be some mechanism 
or parameter that regulates the energy release in GRB afterglows and their decay rate. To satisfy our observations, 
this must occur in such a way that when the energy is released quickly the result is an initially bright afterglow 
which decays quickly. Conversely, if the energy is released slowly over a longer period, the afterglow 
will be less bright initially and decay at a slower rate. This may indicate that there is a narrow range of 
energy provided to the outflow. One possible way to regulate energy release could be continued energy injection. 
If the central engine does not initially release all its energy, but releases it over a much longer period, the 
result could be a fainter afterglow which decays slowly.

\subsection{Observing angle, jet structure and degree of collimation}
The $\rm L_{200\rm{s}}-\alpha_{>200\rm{s}}$ correlation may instead be due to a range in observing angle 
(i.e observer's angle relative to the jet axis), $\theta_{obs}$, with the fainter optical afterglows being 
observed at larger observing angles \citep[e.g][]{ram05,pan08}. \cite{pan08} show in their fig. 3. 
that, for a jet with uniform velocity distribution, an off-axis observer, i.e $\theta_{obs}/\theta_{jet}>1$, 
where $\theta_{jet}$ is the jet opening angle, will observe a shallower decay and observe the afterglow 
to be less luminous in comparison to an observer who is observing closer to the edge of the jet, i.e 
$\theta_{obs}/\theta_{jet}\sim1$. This effect should be observed for both constant density and wind-like 
external media. This model is more complex if structured outflows are considered. For structured outflows, 
off-axis viewers will also observe a shallower and fainter light curve in comparison with on-axis observers, 
but the convergence time and the range of decay rates will vary, depending on how the outflow is structured.

A couple of tests may provide support for the $\rm log\;L_{200\rm{s}}-\alpha_{>200\rm{s}}$ correlation 
resulting from jet structure and observer viewing angle. First, we should expect to see 
convergence of the light curves at late times to a similar decay rate for all observing 
angles. And second, since afterglows that are viewed more off-axis will rise later, we should 
also observe a correlation between afterglow brightness and peak time, though this may be 
complicated by jet structure. \cite{pan08} tested for this correlation using 11 optical 
light curves with rises and find a strong correlation between peak luminosity and 
peak time consistent with this hypothesis though their test could not be applied 
to the GRBs without observed rises \citep[c.f][]{kan10}. 

\vspace{-1mm}
\section{Conclusions}
\label{conclusions}
We computed luminosity light curves at 1600{\AA} for 48 optical/UV GRB afterglows. We find a 
correlation between luminosity at 200s and average decay rate from 200s onwards with 
a significance of 99.998\% (4.2$\sigma$). Regression analysis indicates a linear relationship between decay 
rate and luminosity of $\rm log\;L_{200\rm{s}}=(-3.636\pm0.004)\alpha+(28.08\pm0.13)$. 
 
We used a Monte Carlo simulation to determine, at 4.1$\sigma$ confidence, that the 
$\rm L_{200\rm{s}}-\alpha_{>200\rm{s}}$ correlation is intrinsic and not due to chance or our 
selection criteria. We determined that this correlation is not likely to be a natural consequence 
of the basic synchrotron afterglow model. Instead we find two possible ways to produce the correlation. The first is that 
there is some property of the central engine, outflow or external medium that affects the rate of 
energy release and rate of light curve decay, in such a way that for brighter afterglows the energy is released
more quickly and decays more rapidly than the fainter afterglows. Alternatively, the correlation 
may be produced by a range in observing angles, with observers at large viewing angles witnessing fainter 
and slower decaying light curves. Understanding the origin of this correlation will have important consequences 
on our understanding of the physics and geometry behind GRBs.

\vspace{-1mm}
\section{Acknowledgments}
This research has made use of data obtained from the High Energy Astrophysics Science Archive Research Center 
(HEASARC) and the Leicester Database and Archive Service (LEDAS), provided by NASA's Goddard Space Flight Center 
and the Department of Physics and Astronomy, Leicester University, UK, respectively. SRO, AAB, NPMK and MJP acknowledge 
the support of the UK Space Agency. 

\vspace{-1mm}

\bibliographystyle{mn2e}   
\bibliography{UVOT_LUM_DECAY} 

\IfFileExists{\jobname.bbl}{}
 {\typeout{}
  \typeout{******************************************}
  \typeout{** Please run "bibtex \jobname" to optain}
  \typeout{** the bibliography and then re-run LaTeX}
  \typeout{** twice to fix the references!}
  \typeout{******************************************}
  \typeout{}
 }

\end{document}